\documentclass[aps,twocolumn,superscriptaddress,floatfix,prl]{revtex4}
\usepackage{graphicx,subfigure,amsmath,bbm}

\newcommand{\ua}{\uparrow}
\newcommand{\da}{\downarrow}
\newcommand{\be}{\begin{equation}}
\newcommand{\ee}{\end{equation}}
\newcommand{\bes}{\begin{equation}\begin{split}}
\newcommand{\ees}{\end{split}\end{equation}}

\begin{document}
\title{Nuclear Tuning and Detuning of the Electron Spin Resonance in a Quantum Dot: Theoretical Consideration}
\date{\today}

\author{Jeroen Danon}
\author{Yuli V. Nazarov}
\affiliation{Kavli Institute of NanoScience, Delft University of Technology, 2628 CJ Delft, The Netherlands}
\begin{abstract}
We study nuclear spin dynamics in a quantum dot close to the conditions of electron spin resonance. We show that at small frequency mismatch the nuclear field detunes the resonance. Remarkably, at larger frequency mismatch its effect is opposite: The nuclear system is bistable, and in one of the stable states the field accurately tunes the electron spin splitting to resonance. In this state the nuclear field fluctuations are strongly suppressed and nuclear spin relaxation is accelerated.
\end{abstract}
\maketitle

Electrons confined in semiconductor quantum dots are being investigated intensively in recent years. Much research is inspired by the possibility to use their spin to implement \emph{qubits}, i.e.\ the computational units in a quantum computer~\cite{PhysRevA.57.120}. At present, the main obstacle for this development is the short spin coherence time $T_2^*$ in high-purity quantum dots, measured to be in the ns range~\cite{J.R.Petta09302005,F.H.L.Koppens08262005}. Hyperfine coupling of the electron spin to randomly fluctuating nuclear spins was identified to be the main source of this fast decoherence~\cite{J.R.Petta09302005,klg}.

It was shown recently that hyperfine interaction in semiconductor quantum dots can lead to much richer physics than just dephasing. Various experiments have demonstrated a set of novel phenomena: random switching between two stable states~\cite{F.H.L.Koppens08262005}, current oscillations on a time scale of minutes~\cite{ono:256803}, and strong hysteresis~\cite{tartakovskii:026806}. All new effects were attributed to hyperfine induced dynamical nuclear spin polarization (DNSP) resulting from a non-equilibrium electron spin polarization. The nuclear polarization built up then feeds back to the electron spin splitting and is thereby observed. Optical excitation of quantum dots exhibited  fine mode locking at multiples of the electron spin resonance frequency~\cite{A.Greilich07212006}. The large magnitude of the signal and slow dynamics suggest that DNSP tunes the resonance in individual dots~\cite{dortmund:pc}. Numerical simulations seem to support this point of view, but do not immediately supply a comprehensive picture of the underlying physics~\cite{inarrea:085329}.

These findings triggered ideas to try to make use of the strong feedback~\cite{jouravlev:176804}, and several experiments were designed to optimize the effect of DNSP~\cite{lai:167403,reilly:pc}. It was observed that in a polarized state the fluctuations of the nuclear field are strongly suppressed and their relaxation is accelerated. Both effects may result in a significant improvement of $T_2^*$. On the other hand, controlled DNSP might open up the possibility of processing quantum information in robust nuclear spin ensembles~\cite{nuclei}. All this feeds intensive research on the coupled electron-nuclear spin dynamics in quantum dots~\cite{petta:rudner}.

Recently, electron spin resonance (ESR) in a double quantum dot~\cite{frank:nature}, has been used to perform single electron spin rotations. Besides the demonstration of spin rotations, the experiment gave a clear indication of DNSP. The resonant response extended to a rather broad frequency interval without any amplitude reduction. Remarkably, this broadening was asymmetric with respect to the ESR frequency $g\mu_B B_0/\hbar$. Besides, a significant hysteresis was observed: The response depended on the sweep direction of frequency or magnetic field, suggesting that the resonant condition is shifted during the sweep, as if something tunes the electron splitting. While these effects have been speculatively attributed to DNSP, their exact mechanisms are not yet understood.

\begin{figure}[b]
\includegraphics[width=8.5cm]{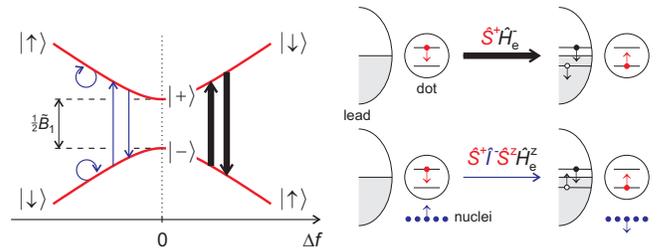}
\caption{A quantum dot under ESR conditions. Left: The spin-split levels $|+\rangle$ and $|-\rangle$ in the rotating wave approximation. The arrows show the \emph{first-order} transitions with a dissipative electron spin-flip (thick) and \emph{second-order} transitions with a nuclear spin-flip (thin). Right: The initial and final states of the transitions for the case when dissipation is dominated by co-tunneling to the leads.}\label{fig:levels}
\end{figure}
In this work we present a simple model to study the coupled electron-nuclear spin dynamics in a \emph{single} quantum dot close to the ESR condition, assuming that the ac driving is sufficiently strong to saturate the resonance. We find that ESR polarizes the nuclei in a preferred direction. We show that this results in tuning as well as detuning of the resonance by DNSP, depending on the mismatch between the driving frequency and $g\mu_B |B_0|/\hbar$: At small mismatch, the nuclear field built up simply detunes the spin splitting away from ESR. At larger mismatch, competition between ESR pumping and nuclear spin relaxation causes a bistability, and in one of the stable states the nuclear field actually tunes the system back to ESR. With this model, we find a number of recently observed effects (strong asymmetric hysteresis~\cite{frank:nature,tartakovskii:026806} and a reduction of nuclear field fluctuations and accelerated dynamics~\cite{reilly:pc}), and we provide a clear explanation of the mechanisms involved. To achieve a quantitative agreement with the experiments, one would have to use more detailed and specific models to account for, e.g.\ the presence of two coupled dots~\cite{frank:nature} or the possibility of substantial electric contributions to the ESR signal~\cite{laird:0707.0557}.

In our model, a single electron is confined in a quantum dot with its energy well below the Fermi levels of the nearby leads. To achieve ESR, one combines a dc and ac magnetic field, $\mathbf{B}_{\text{tot}} = B_0\hat z + B_1 \cos (\omega t) \hat x$, the ESR condition being $\hbar\omega = g\mu_B|B_0|$. The interaction of the total magnetic field and the electron spin $\mathbf{\hat S}$ is given by the Hamiltonian $\hat H = -g \mu_B \mathbf{B}_\text{tot} \cdot \mathbf{\hat S}$.

It is natural to assume that the ESR frequency mismatch $\Delta f \equiv \omega-g\mu_B|B_0|/\hbar$ as well as the Rabi frequency $\tilde{B}_1 \equiv g\mu_BB_1/\hbar$ are much smaller than $\omega$. This justifies a rotating wave approximation. In a rotating frame, we can write the Hamiltonian as
\begin{equation}\label{eq:h0}
 \hat H_0 = \hbar(\Delta f) \hat S^z + \frac{\hbar}{2}\tilde{B}_1 \hat S^x.
\end{equation}
This $\hat H_0$ determines the effective electron spin eigenstates in the rotating frame, $|+\rangle = \cos \frac{1}{2} \theta |\!\ua\rangle + \sin \frac{1}{2} \theta |\!\da\rangle$ and $|-\rangle = \sin \frac{1}{2} \theta |\!\ua\rangle -\cos \frac{1}{2} \theta |\!\da\rangle$,
where $\theta = \pi/2 +\arctan(2 \Delta f/\tilde{B}_1)$. To determine the probabilities $\rho_{\pm}$ to be in one of the eigenstates, we have to take into account dissipative processes accompanied by a spin-flip. Those are due to the coupling to the environment, that very generally can be represented as
\begin{equation}
\label{eq:env}
\hat H_\text{coup} = {\hat H}^z_e {\hat S}^z +\frac{1}{2} \left\{ {\hat H}^{-}_e {\hat S}^{+} e^{i\omega t} +
{\hat H}^{+}_e {\hat S}^{-}e^{-i\omega t} \right\},
\end{equation}
where $\hat{H}^{z,\pm}_e$ represent the environmental degrees of freedom coupled to corresponding electron spin components. It is convenient to consider the environment in the rest frame, while $\mathbf{\hat S}$ is defined in the rotating frame. This makes the coupling explicitly time-dependent. For quantum dots, relevant dissipation mechanisms are: (i) electron-hole pair creation in a nearby lead by a co-tunneling process \cite{dreiser:0705.3557}, (ii) spin-orbit interaction with phonons \cite{PhysRevB.64.125316}, (iii) direct coupling to phonons perturbed by Zeeman energy. Mechanisms (i) and (ii) can lead to an electron spin-flip in the $z$-direction in first order, not causing any nuclear spin flips. Mechanism (iii) couples only to $S_z$ and therefore it can flip the electron spin only through a second-order process involving the nuclei~\cite{PhysRevB.66.155327}.

We assume temperatures lower than $\hbar\omega$. We then find from energy consideration that the terms proportional to $\hat {H}_e^- \hat S^+e^{i\omega t}$ dominate the transition rates between $|+\rangle$ and $|-\rangle$, since they correspond to the maximum energy transfer $\approx g\mu_B|B_0| \approx \hbar \omega$ \emph{from} the dot \emph{to} the environment (see Fig.\ \ref{fig:levels}). To find the steady state probabilities $\rho_+$ and $\rho_-$, we compute these rates and solve the master equation $ \Gamma^{-}_{r}[\omega]\left(-M_{-+} \rho_+ +M_{+-} \rho_-\right) = 0,$ where $\Gamma^{-}_{r}$ is the maximum relaxation rate, $\Gamma^{-}_r [\omega ] = \int  \langle \hat{H}^-_e(0)  \hat{H}^+_e(\tau) \rangle e^{-i\omega\tau} d\tau/4\hbar^2$. For mechanism (i) $\Gamma^{-}_r \propto \omega$, while for (ii) $\Gamma^{-}_r \propto \omega^{5}$. The matrix elements $M_{\alpha\beta} \equiv | \langle \alpha|\hat S^+|\beta\rangle|^2$ are calculated from $|+\rangle$ and $|-\rangle$, and this yields $\rho_\pm = \frac{1}{2} \pm \cos \theta / (1+\cos^2 \theta)$. Far from the resonance (if $\theta \to 0,\pi$) the spin is in the ground state $|\!\!\uparrow\rangle$ (corresponding with $\rho_+ \to 1$ or $\rho_- \to 1$, see Fig. \ref{fig:levels}), while exactly at resonance ($\theta=\pi/2$) one finds $\rho_\pm =1/2$. This approach is valid provided that the energy splitting in the rotating frame is sufficiently big, i.e.\ $\tilde{B}_1 \gg \Gamma_{r}^-[\omega]$.

Let us now consider the nuclear spins $\mathbf{\hat I}$, which are coupled to the electron spin via hyperfine interaction~\cite{hyperfinerev},
\begin{equation}
\hat H_{\text{hf}} = \frac{E_n}{2N} \sum_k \left\{ 2\hat S^z \hat I^z_k + \hat S^+ e^{i\omega t} \hat I^-_k + \hat S^-e^{-i\omega t}\hat I^+_k \right\},
\end{equation}
where the sum runs over all nuclei. For simplicity we assume that all nuclear spins are equally strongly coupled to the electron spin, so that the prefactor reduces to the hyperfine coupling energy (for bulk GaAs $IE_n \sim$ 100 $\mu eV$~\cite{hyperfinerev}) divided by the effective number of nuclei $N$. For quantum dots, this number is big ($N \sim$ 10$^6$), and this distinguishes the situation in quantum dots from that of a single paramagnetic ion~\cite{PhysRev.98.1729}.

The effect of the hyperfine interaction is twofold. Firstly, the nuclei affect the electron dynamics via the Overhauser field $\langle I^z\rangle E_n$, that adds to the static $z$-component of the external magnetic field. Secondly, the interaction can cause electron spin flips to be accompanied by flips of nuclear spins. If there is a preferential direction for these flips, they can be the source of DNSP.

Let us evaluate the rate of the hyperfine induced nuclear spin flips. We keep in mind that hyperfine interaction by itself cannot cause spin exchange between the electron and the nuclear system owing to the energy mismatch $\approx \hbar \omega$ between the initial and final state. The rate thus arises from a second-order process, the corresponding amplitude incorporating $\hat{H}_\text{coup}$ and $\hat{H}_{\text{hf}}$. In principle, there are six processes capable of flipping nuclear spins. To estimate their relative magnitude, we note that the environment favors large positive energy absorption. This brings us to the conclusion that the dominant process comes from combination of $\hat S^+e^{i\omega t} \hat I^-$ in $\hat{H}_{\text{hf}}$, and $\hat S^z\hat H_e^z$ in $\hat{H}_\text{coup}$ (see Fig.\ \ref{fig:levels}). The corresponding energy transfer is $\approx \hbar\omega$. This means that nuclear spins {\it only} flip from the ``up" to the ``down" state: There is a preferential direction, needed for DNSP~\cite{endnote1}. The resulting polarization is negative, $P\equiv (N_\uparrow - N_\downarrow)/N <0$, $N_{\uparrow(\downarrow)}$ being the number of nuclei with spin ``up"(``down").

The pumping rate is proportional to $\Gamma^{z}_r[\omega ] = \int  \langle \hat{H}^z_e(0)  \hat{H}^z_e(\tau) \rangle e^{-i\omega\tau} d\tau/\hbar^2$, accounting for the dissipation of $\hbar \omega$. For mechanism (i), $\Gamma^-_{r}= \Gamma^z_{r}$ owing to SU(2) symmetry in spin space. For spin-orbit mechanism (ii), $\Gamma^-_{r}$ and $\Gamma^z_{r}$ are of the same order of magnitude~\cite{PhysRevB.64.125316}. The total pumping rate reads
\begin{equation}\label{eq:rate1}
\Gamma_p = -\frac{\Gamma^z_{r}[ \omega ]E^2_n}{4N^2(\hbar \omega)^2} N_\uparrow \sum_{k,l\, \in \,\{+,-\}} M_{kl}\rho_l .
\end{equation}
We assume $P \ll 1$,  so that $N_\uparrow = N/2$ (even small polarizations are enough to (de)tune the resonance in a wide frequency range). At this stage we incorporate the effect of the Overhauser field: It is a simple shift of the frequency mismatch, and we define the resulting mismatch $\Delta =\Delta f + IE_nP/\hbar$. We note that the validity of the rotating wave approximation now requires $|\Delta|\ll \omega$. The matrix elements in (\ref{eq:rate1}) and $\rho_{\pm}$ are now functions of $\Delta$ and the pumping rate assumes a Lorentzian shape
\begin{equation}\label{eq:lor}
\Gamma_p (\Delta) = -\frac{5\Gamma_r^z[\omega] E_n^2}{32 N(\hbar \omega)^2} 
\frac{1}{1 + 8(\Delta/\tilde{B}_1)^{2}},
\end{equation}
with the same width as e.g.\ $\langle S_z (\Delta) \rangle$. The numerical factor accounts for $I=3/2$ for GaAs. We see that $\Gamma_p \ll \Gamma_r$, provided the Zeeman splitting $\hbar \omega$ exceeds the typical fluctuations of the nuclear field $\hbar \Omega = IE_n/\sqrt{N}$. This sets the limits of validity of our perturbative approach.
 
The resulting nuclear polarization follows from the competition between spin pumping and intrinsic nuclear spin relaxation characterized by the rate $1/\tau_n$. In terms of $P$, the balance equation reads
\begin{equation}\label{eq:rate2}
\frac{dP}{dt} = \frac{2 \Gamma_p(\Delta)}{N} -\frac{1}{\tau_n} P,
\end{equation}
which is in combination with the Lorentzian in Eq.\ (\ref{eq:lor}) the main result of our work.

To proceed, let us note that the nuclear relaxation rate is very low, $\tau_n \sim 10$~s~\cite{paget:prb}. We express this smallness in terms of a dimensionless parameter $\alpha = (18/5\sqrt2) (\tau_n \Gamma^{z}_r[\omega ])^{-1} \sqrt{N} (\tilde{B}_1\omega^2/\Omega^3)$. As $\alpha \propto \tilde B_1 / \Gamma_p(0) \tau_n$, a small $\alpha$ means that $\Gamma_p$ is sharply peaked compared to the slow relaxation rate $1/\tau_n$. Although the theory outlined is valid for any $\alpha$, a strong DNSP feedback requires $\alpha \ll 1$.  We assume this from now on.

Let us consider {\it detuning} first. The natural measure for the frequency mismatch is the resonance width $\tilde{B}_1/\sqrt{2}$. If the initial frequency mismatch is small, $\Delta f/\tilde{B}_1 \lesssim 1$, then the weak relaxation stops the building of nuclear polarization only at significantly large $\Delta/\tilde{B}_1= 2^{-3/2}\alpha^{-1/3}\gg 1$, that is, far from the resonance (see Fig.\ 2a).

Counterintuitively, a larger frequency mismatch results in {\it tuning}. If $\Delta f/\tilde{B}_1 > 3\cdot 2^{-13/6} \alpha^{-1/3}$, then Eq.\ (\ref{eq:rate2}) has three zeros, and the polarization becomes bistable, as in Fig.\ 2b. This bistability is preserved till much bigger mismatches, with upper boundary $(\Delta f)_{\text{max}} \equiv \tilde{B}_1 2^{-3/2} \alpha^{-1}$. In one of the stable configurations $\Delta \simeq \tilde{B}_1$, and the system is tuned to resonance. The other stable state is unpolarized and thus far away from the ESR condition. We stress that the bistability
is asymmetric: If $E_n >0$, as in GaAs (an antiparallel arrangement of electron and nuclear spins is energetically favorable), the bistability occurs \emph{only} at $\Delta f >0$. If on the other hand $E_n <0$, it occurs only at $\omega$ lower than $g\mu_B|B_0|/\hbar$. 

\begin{figure}[t]
\includegraphics[scale=0.8]{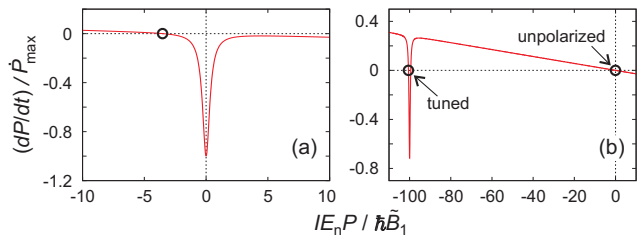}
\caption{Two plots of $dP/dt$ normalized by the maximum ESR pumping rate $\dot P_\text{max} = 2\Gamma_p(0)/N$, versus the nuclear polarization $P$. The circles indicate the stable configurations. (a) Detuning at small frequency mismatch ($\Delta f = 0$). (b) Bistability and tuning at large frequency mismatch ($\Delta f = 100\tilde B_1$). For both plots $\alpha = 10^{-3}$.}\label{fig:dpdt}
\end{figure}
Such a bistability implies also hysteretic behavior. Let us adiabatically sweep the frequency starting at $\omega<g\mu_B|B_0|$ far from the resonance (see Fig.\ 3). Upon increasing $\omega$ we first cross $g\mu_B|B_0|$ (i.e.\ the line $\Delta f = 0$), and then get to the tuned state at $\Delta f \simeq \tilde{B}_1\alpha^{-1/3}$. We remain in this state until the frequency mismatch reaches $(\Delta f)_{\text{max}}$, provided that our sweep speed is much smaller than the typical nuclear spin pumping rate, i.e.\ $\dot \omega \ll 2E_n\Gamma_p(0)/N\hbar$. If we cross $(\Delta f)_{\text{max}}$, nuclear relaxation becomes stronger than DNSP: The tuning ceases and the only stable state is again the unpolarized one, which will be reached on a time scale of $\tau_n$. If we then go backwards decreasing $\omega$, we will not get into the tuned state but remain in the stable unpolarized state.
 
The overall structure in the ($B_0,\omega$)-plane is sketched in Fig.\ 3, where the bistability occurs in the gray-shaded region. An experimentally accessible quantity is the width of this region. It is equal to the maximum frequency mismatch $(\Delta f )_{\text{max}} \propto \alpha^{-1} \propto \Gamma_r^z [\omega ] / \omega^2$, so it exhibits dependence on $\omega$. At lower frequencies, mechanism (i) (interaction with the leads) dominates the dissipative spin-flips and $\Gamma_r^z [\omega]\propto\omega$, so that $(\Delta f )_{\text{max}}\propto 1/\omega$. At larger frequencies mechanism (ii) takes over, resulting in $(\Delta f )_{\text{max}}\propto \omega^3$. This together implies that $(\Delta f)_{\text{max}}$ reaches a minimum at a finite $\omega_c$. We illustrate this behavior in Fig.\ 3, choosing $(\Delta f)_\text{max} = 0.04\,\omega_c [ 3(\omega_c / \omega) + ( \omega / \omega_c)^3]$.
The separate contributions of mechanism (i) and (ii) are indicated with thin dashed lines.

\begin{figure}[t]
\includegraphics[scale=.65]{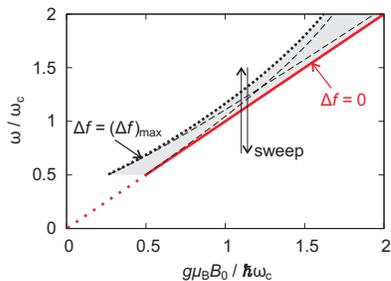}
\caption{Stability diagram in the ($\omega , B_0$)-plane for ESR induced DNSP. Bistability and tuning to the resonance occur in the gray-shaded region. The arrows show an adiabatic frequency sweep leading to the tuned state. The dashed lines present the separate contributions of mechanism (i) and (ii). This diagram qualitatively agrees with~\cite{frank:nature}, Fig.\ 2C} \label{fig:stab}
\end{figure}
Let us give a numerical example supporting the assumptions made. Based on typical experimental parameters~\cite{frank:nature}, we use $\tau_n = 15$~s, $N=10^6$, $\tilde{B}_1 = 1.5$~mT, $\omega = 120$~mT and $\Omega = 5$~mT (for $|g|=0.35$~\cite{frank:nature}, 1~mT $\approx 3\cdot 10^7$~s$^{-1}$). We take $\Gamma^z_r = 2\cdot 10^6$~s$^{-1}$, this is in accordance with a lower bound estimate set by the typical leakage current of 100 fA~\cite{frank:nature}. We find that $\alpha \approx 1.5 \cdot 10^{-2}$, so it is small indeed, and this suggests strong DNSP. The same set of parameters gives $(\Delta f)_\text{max} \approx 24\tilde B_1$, which is much bigger than the resonance width and even comparable to the driving frequency $(\Delta f)_\text{max} \approx 0.3 \, \omega$. This gives 40~mT, a value that agrees well with experimental observations in a double dot (\cite{frank:nature}, Fig.\ 2C). The maximum polarization is achieved at the edge of the region, where $P_\text{max} \approx 7.2\cdot 10^{-3}$, so that $P \ll 1$ indeed.

We also investigated both the switching rates between the tuned and the unpolarized state, and the small fluctuations near these stable states. To estimate the fluctuations, we use a Fokker-Planck equation for the distribution function of the polarization ${\cal P}(P)$. To derive the equation, we regard the nuclear dynamics as a random walk on a discrete set of spin values $n=\frac{1}{2}(N_\uparrow-N_\downarrow)$. The pumping rate $\Gamma_p$ only causes transitions from $n$ to $n-1$, while the spin relaxation causes transitions in both directions with almost equal rates $(1/2\tau_n) N_{\uparrow,\downarrow}\gg \Gamma_p$. We go to the continuous limit, justified by 
the large number of nuclei ($\sim 10^6$) to obtain~\cite{vankampen}
\begin{equation}\label{eq:fp}
\frac{\partial \mathcal{P}}{\partial t} = 
\frac{\partial}{\partial P} 
\left\{ \mathcal{P} \left[\frac{1}{\tau_n} P + \frac{2}{N}\Gamma_p \right] + 
\frac{\partial}{\partial P}\mathcal{P}\frac{1}{N\tau_n}\right\}.
\end{equation}
From the steady state solution of (\ref{eq:fp}) we evaluate the small fluctuations around the unpolarized and tuned states. While $\left\langle (\Delta P)^2\right\rangle_\text{unp} = 1/N$ is not affected by ESR, the fluctuations in the tuned state are suppressed roughly by a factor $\alpha^{-1}$ (67 for the numerical example), more precisely $\left\langle (\Delta P)^2\right\rangle_\text{tun}= \frac{1}{2} \alpha \{ q^3(1 - q)\}^{-1/2}\left\langle (\Delta P)^2\right\rangle_\text{unp},$ where $q\equiv \Delta f / (\Delta f)_\text{max} \sim 1$. Importantly, this factor also determines the {\it acceleration} of the nuclear dynamics: The local nuclear spin relaxation time in the tuned state, $\tau^\text{tun}_n$, is shorter than $\tau_n$ by the same factor.

This quenching of the fluctuations is also a justification for neglecting them. If the fluctuations would have been fully developed, one could only neglect them if the resonance width $\tilde{B}_1 \gg \Omega$. Since the fluctuations are suppressed, this condition is now achieved at much smaller driving fields $\tilde{B}_1 \gg \Omega\sqrt{\alpha}$. The same condition guarantees that spontaneous switching between the tuned and unpolarized state
occurs with an exponentially small rate. We evaluate this rate from (\ref{eq:fp}) with Kramers method \cite{vankampen} to obtain
\begin{equation}
\Gamma_{t\to u} = \frac{1}{2\pi\tau^\text{tun}_n} \exp\left( -\frac{\tilde{B}^2_1}{ 
4\Omega^2 {\alpha}} f(q) \right),
\end{equation}
where $f(q) =\arctan \sqrt{q^{-1}-1} -\sqrt{q(1-q)} \simeq 1$. The values from our numerical example give $1/\Gamma_{t\to u} \simeq 4$~s at $q=\frac{1}{2}$. The inverse rate $\Gamma_{u\to t}$ is yet smaller, so that if the dot has switched to the unpolarized state, it is unlikely to switch back by itself. One would have to make again a frequency sweep as described above.

To conclude, we have investigated DNSP under ESR conditions to find both detuning and bistability-related tuning of the resonance. The simple model in use explains qualitatively a set of recent experimental findings. The authors gladly acknowledge useful communications with F.H.L. Koppens, L.M.K. Vandersypen, M.S. Rudner, L.S. Levitov, D.R. Yakovlev and D.J. Reilly. This work was supported by the Dutch Foundation for Fundamental Research on Matter (FOM).


\begin{thebibliography}{99}

\bibitem{PhysRevA.57.120} D. Loss and D. P. Divincenzo, Phys. Rev. A \textbf{57}, 120 (1998).

\bibitem{J.R.Petta09302005} A. C. Johnson, et al., Nature \textbf{435}, 925 (2005).\\ J. R. Petta, et al., Science \textbf{309}, 2180 (2005).

\bibitem{F.H.L.Koppens08262005} F. H. L. Koppens, et al., Science \textbf{309}, 1346 (2005).

\bibitem{klg} A. V. Khaetskii, et al., Phys. Rev. Lett. \textbf{88}, 186802 (2002).

\bibitem{ono:256803} K. Ono and S. Tarucha, Phys. Rev. Lett. \textbf{92}, 256803 (2004).

\bibitem{tartakovskii:026806} A. I. Tartakovskii, et al., Phys. Rev. Lett. \textbf{98}, 026806 (2007).

\bibitem{A.Greilich07212006} A. Greilich, et al., Science \textbf{313}, 341 (2006).

\bibitem{dortmund:pc} A. Greilich, et al., Science \textbf{317}, 1896 (2007).

\bibitem{inarrea:085329} J. I\~{n}arrea, et al., Phys. Rev. B \textbf{76}, 085329 (2007).

\bibitem{jouravlev:176804} O. N. Jouravlev and Yu. V. Nazarov, Phys. Rev. Lett. \textbf{96}, 176804 (2006).

\bibitem{lai:167403} C. W. Lai, et al., Phys. Rev. Lett. \textbf{96}, 167403 (2006).

\bibitem{reilly:pc} D. J. Reilly, et al., in preparation (2007).

\bibitem{nuclei} J. M. Taylor, et al., Phys. Rev. Lett. \textbf{90}, 206803 (2003).

\bibitem{petta:rudner} J. R. Petta, et al., arXiv:0709.0920 (2007). M. S. Rudner and L. S. Levitov, Phys. Rev. Lett. \textbf{99}, 036602 (2007).

\bibitem{frank:nature} F. H. L. Koppens, et al., Nature \textbf{442}, 766 (2006).

\bibitem{laird:0707.0557} E. A. Laird, et al., arXiv:0707.0557 (2007).

\bibitem{dreiser:0705.3557} J. Dreiser, et al., arXiv:0705.3557 (2007).\\ A. Kaminski, et al., Phys. Rev. B \textbf{62}, 8154 (2000).

\bibitem{PhysRevB.64.125316} A. V. Khaetskii and Yu. V. Nazarov, Phys. Rev. B \textbf{64}, 125316 (2001).

\bibitem{PhysRevB.66.155327} S. I. Erlingsson and Yu. V. Nazarov, Phys. Rev. B \textbf{66}, 155327 (2002).

\bibitem{hyperfinerev} D. Paget, et al., Phys. Rev. B \textbf{15}, 5780 (1977).

\bibitem{PhysRev.98.1729} A. Abragam, Phys. Rev. \textbf{98}, 1729 (1955).

\bibitem{endnote1} Although $I=3/2$ in GaAs, this only gives rise to a difference in numerical prefactors.

\bibitem{paget:prb} D. Paget, Phys. Rev. B \textbf{25}, 4444 (1982).

\bibitem{vankampen} N. G. van Kampen, \textit{Stochastic Processes in Physics and Chemistry} (North-Holland, Amsterdam, 1990).

\end{thebibliography}
\end{document}